\documentclass[floatfix,aps,twocolumn,superscriptaddress]{revtex4-1}

\usepackage{amssymb}
\usepackage{amsmath}
\usepackage[dvips]{graphicx}
\usepackage{bm}
\usepackage{float}
\usepackage{color}

\newcommand{\be}{\begin{equation}}
\newcommand{\ee}{\end{equation}}
\newcommand{\bea}{\begin{eqnarray}}
\newcommand{\eea}{\end{eqnarray}}

\begin{document}

\title{Spin filtering and thermopower in star coupled quantum dot devices}

\author{J. A. Andrade}
\affiliation{Centro At{\'o}mico Bariloche and Instituto Balseiro, CNEA, 8400 Bariloche, Argentina}
\affiliation{Consejo Nacional de Investigaciones Cient\'{\i}ficas y T\'ecnicas (CONICET), Argentina}
\author{Pablo S. Cornaglia}
\affiliation{Centro At{\'o}mico Bariloche and Instituto Balseiro, CNEA, 8400 Bariloche, Argentina}
\affiliation{Consejo Nacional de Investigaciones Cient\'{\i}ficas y T\'ecnicas (CONICET), Argentina}

\begin{abstract}
We analyze the linear thermoelectric transport properties of devices with three quantum dots in a star configuration. A central quantum dot is tunnel-coupled to source and drain electrodes and to two additional quantum dots.  For a wide range of parameters, in the absence of an external magnetic field, the system is a singular Fermi liquid with a non-analytic behavior of the electric transport properties at low energies. The singular behavior is associated with the development of a ferromagnetic or an underscreened Kondo effect, depending on the parameter regime. 
A magnetic field drives the system into a regular Fermi liquid regime and leads to a large peak ($\sim k_B/|e|$) in the spin thermopower as a function of the temperature, and to a $\sim 100\%$ spin polarized current for a wide range of parameters due to interference effects. We find a qualitatively equivalent behavior for systems with a larger number of side coupled quantum dots, with the maximum value of the spin thermopower decreasing as the number of side-coupled quantum dots increases.

\end{abstract}

\maketitle
The use of the electron spin degree of freedom for classical or quantum computing operations and for information storage and transmission is what is usually called spintronics~\cite{wolf2001spintronics,vzutic2004spintronics}. The main advantages of a spin based electronics are a reduced dissipation and a faster operation speed~\cite{kent2015new}.
A key step towards the implementation of spintronic devices is the ability to generate and inject spin currents in semiconductor based components~\cite{datta1990electronic,hatami2007thermal,wolf2001spintronics,vzutic2004spintronics}. Spin currents can be injected using ferromagnetic contacts or generated directly in the non-magnetic material by applying external magnetic fields.

There have been several proposals to generate spin polarized currents using quantum dots (QDs) built in semiconducting heterostructures. These proposals generally involve small QDs where a single electronic level is relevant for the transport properties at low energies \cite{recher2000quantum,potok2003polarized}. In the regime where the electronic level is singly occupied, an external magnetic field can polarize the spin of the electron in the QD level producing a spin dependent transmittance through it. To obtain a large spin polarization of the current, however, large magnetic fields and a fine tunning of the level energy are usually needed.
The energy associated with the Zeeman splitting needs to be of the order of the level hybridization and the QD level to be near a resonance condition with the Fermi energy of the electrodes. 
Other proposals in multiple QD devices involve using interference effects to generate fully spin polarized currents, applying an external magnetic field such that there is a destructive interference for one of the spin components leading to a vanishing transmittance \cite{torio2004spin,song2003fano,ojeda2009array}.

The measurement of the thermoelectric effect in molecular junctions \cite{reddy2007thermoelectricity} and the observation of the spin Seebeck effect in magnetic conductors \cite{uchida2008observation} spawned numerous studies on the thermal generation of charge and spin currents in nanoscopic devices \cite{konig2008,murphy2008optimal,finch2009giant,Costi2010,Weymann,dubi2011colloquium,Cornaglia2012,widawsky2012,rejec2012,rourabas2012}. The spin Seebeck effect could be used to generate pure spin currents in molecular or QD devices, i.e. without having an associated charge current \cite{rejec2012}. 
It was early recognized that sharp features in the electronic density of states near the Fermi energy can lead to an enhancement of thermopower \cite{mahan1996best}. 
In QDs, the development of the Kondo effect at temperatures below a characteristic scale $T_K$ generates a narrow peak, so-called Kondo peak or Abrikosov-Suhl resonance, of width $\sim k_BT_K$ at the Fermi energy in the spectral density of the QD. The regular Kondo effect involves a spin 1/2 coupled antiferromagnetically to the conduction electron band and leads to an Abrikosov-Suhl resonance centered near the Fermi level (the shift is $\ll k_BT_K$)  with a low electron-hole asymmetry and a small thermoelectric response at low temperatures~\cite{Costi2010}. Other varieties of the Kondo effect with a higher impurity symmetry, as the SU(4) case observed in carbon nanotube junctions, lead to an Abrikosov-Suhl resonance shifted by $\sim k_BT_K$ above the Fermi energy and to a large charge thermoelectric response~\cite{rourabas2012,lim2014orbital}. To obtain a large spin thermoelectric response in these Kondo systems, however, a Zeeman splitting larger than the Kondo scale $k_BT_K$ is needed.

The underscreened Kondo effect, which has been observed in molecular devices~\cite{roch2008quantum,roch2008quantum,Parks11062010}, and the ferromagnetic Kondo effect have been predicted to occur in multiple quantum dot devices ~\cite{kuzmenko2006,mitchell2009,mitchell2013,Baruselli2013,andrade2015ferro,Tooski2016366}. An appealing property of these Kondo systems is that they lead to a singular Fermi liquid behavior~\cite{varma2002singular,mehta2005regular,coleman2003pepin,PhysRevLett.94.036802,koller2005singular} with a logarithmic dependence on the excitation energy of the low energy properties. In these devices a spin decouples asymptotically from the electron bath, at low temperatures, and becomes easily polarizable with any finite external magnetic field, leading to high magnetotransport and spin thermoelectric responses \cite{Cornaglia2012,zitko2013,cabrera2013magneto}.

We study the thermoelectric properties of a device with three quantum dots in a star configuration which leads to a realization of the underscreened Kondo effect for a spin $S=1$ and to the ferromagnetic Kondo effect for a spin $S=1/2$ \cite{kuzmenko2006,mitchell2009,mitchell2013,Baruselli2013,andrade2015ferro,Tooski2016366}. We show that these devices can be used as spin filters and to thermally generate spin polarized currents. We find that in the ferromagnetic Kondo regime the singular properties emerge at a scale much larger than in the underscreened Kondo regime, making it more suitable for the experimental observation of these effects. 

The  rest of the paper is organized as follows. In Sec. \ref{sec:model} we present the model and summarize previous results on the low energy properties of the system. In Sec. \ref{sec:transport} we present numerical results, using the Full Density Matrix extension \cite{Andreas2007} of Wilson's Numerical Renormalization Group (FDM-NRG) \cite{Wilson1975kondo,bulla2008numerical}, for the conductance, thermoelectric power, magnetoconductance and spectral density as a function of temperature and gate voltage. Finally, in Sec. \ref{sec:sumconcl} we summarize the main results and present the conclusions.
\section{Model}\label{sec:model}
\begin{figure}[t]
\includegraphics[width=0.8\columnwidth]{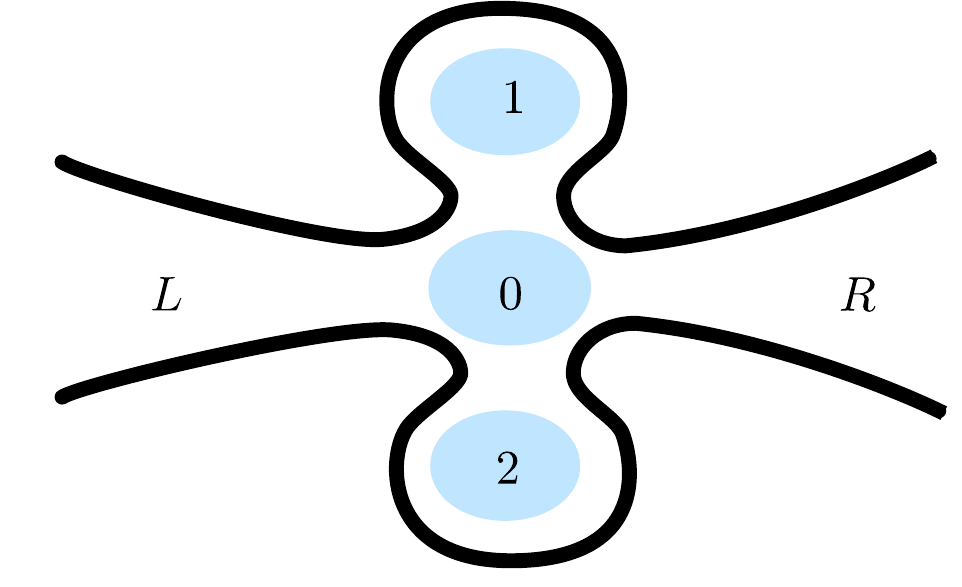}
\caption{(Color online) Schematic representation of a three QD device in a star configuration. The central QD which is labeled with $0$ is tunnel coupled to two QDs and to left (L) and right (R) electrodes.}
\label{fig:device}
\end{figure}

We analyze a model device with three QDs tunnel coupled in a star configuration (see Fig. \ref{fig:device}). A central QD is coupled to two QDs and to left and right electrodes.  The device is described by the following Hamiltonian \cite{andrade2015ferro}

\begin{eqnarray} \label{eq:hamilt}
	H&=&\sum_{\ell=0}^{2}\left(\sum_{\sigma=\uparrow,\downarrow}\varepsilon_{\ell\sigma}\hat{n}_{\ell\sigma}+ \frac{U_\ell}{2} \hat{n}_{\ell\uparrow}\hat{n}_{\ell\downarrow}\right) \nonumber\\
&& +\sum_{\sigma}\sum_{\ell=1}^{2}\left(t_{\ell}d_{\ell\sigma}^{\dagger}d_{0\sigma}+h.c.\right) \nonumber\\
&& + \sum_{\nu=L,R}\sum_{k,\sigma}\left(V_{k\nu}d_{0\sigma}^{\dagger}c_{\nu k\sigma}+h.c.\right) \nonumber\\
&& +\sum_{\nu, k,\sigma} \epsilon_{k\nu} c_{\nu k\sigma}^\dagger c_{\nu k\sigma},
\end{eqnarray}
where we consider a single electronic orbital on each QD \footnote{For a small enough quantum dot, the linear transport properties are governed by its lowest unoccupied or the highest occupied orbital.} with energy $\varepsilon_{\ell\sigma}$, $\hat{n}_{\ell\sigma}= d_{\ell\sigma}^\dagger d_{\ell\sigma}$ is the electron number operator of the $\ell$-th QD, $U_\ell$ is its charging energy, and $\sigma$ is the electron spin projection along the $\hat{z}$ axis. An external magnetic field $B\hat{z}$ produces a Zeeman splitting of the level energy of each orbital as $\varepsilon_{\ell\uparrow}=\varepsilon_\ell- g\mu_B B/2$ and $\varepsilon_{\ell\downarrow}=\varepsilon_\ell + g\mu_B B/2$. The energies $\varepsilon_{\ell}=U_{\ell}-C_{g\ell} V_{g\ell}$ can be experimentally controlled via gate voltages $V_{g\ell}$. 
The first term in Eq. (\ref{eq:hamilt}) describes the electrostatic interaction on the QDs. 
The second term describes the tunneling coupling between the central QD ($\ell=0$) and the two side-coupled QDs. The third term 
describes the coupling between the central QD and the left ($L$) and right ($R$) electrodes, 
which are modeled by two non-interacting Fermi gases. 

The low energy properties of this model, for $N$ side-coupled QDs, were analyzed for $B=0$ in Ref. \cite{andrade2015ferro} in the Kondo regime with single electron per QD. When the tunnel coupling between the side coupled QDs and the central QD is much larger than the hybridization between  the central QD and the leads, the low energy physics is governed by a ferromagnetic Kondo Hamiltonian for a $S = (N-1)/2$ impurity spin.
In weak interdot coupling regime, a two stage Kondo effect develops as the temperature is decreased. The first stage is due to the Kondo screening of the spin on the central QD by the leads, and the second one at lower temperatures is an underscreened 
Kondo effect between a local heavy Fermi liquid at the central QD and the spins on the side coupled QDs. 
The heavy quasiparticles at the central QD can only partially screen the magnetic moment on the side coupled QDs leaving at low temperatures a residual spin with a ferromagnetic coupling to the quasiparticles. As in the case of ferromagnetic Kondo effect, the underscreened Kondo effect presents singular liquid Fermi properties due to a logarithmic decoupling at low energy of residual spin.

For simplicity we consider below $t_\ell=t$ and $U_\ell=U$. The main conclusions, however, do not depend on the homogeneity of the QDs' couplings and charging energies.

\section{Transport properties}\label{sec:transport}
We analyze the transport properties in the linear response regime for both the bias voltage $\Delta V=(\mu_L-\mu_R)/e$ and the temperature difference $\Delta T=T_L-T_R$ between left and right electrodes. The current for the spin projection $\sigma$ can be written as \cite{Costi2010}
\begin{equation}
I_{\sigma} = G_\sigma \Delta V +G_\sigma S_{\sigma} \Delta T,
	\label{eq:linres}
\end{equation}
where the conductance $G_\sigma$ and the Seebeck coefficient $S_{\sigma}$ are given
by
\begin{equation}
G_\sigma=\frac{e^{2}}{h}\,\mathcal{I}_{0\sigma}\,,\qquad S_{\sigma}=-\frac{k_{B}}{|e|}\frac{%
\mathcal{I}_{1 \sigma}}{k_{B}T\,\mathcal{I}_{0\sigma}},  \label{GandS}
\end{equation}
with $\mathcal{I}_{n\sigma }=\int_{-\infty }^{\infty }\omega ^{n}\left( -\frac{%
\partial f(\omega )}{\partial \omega }\right) \mathcal{T}_{\sigma }(\omega
)\,\mathrm{d}\omega \,$. Here $\mathcal{T}_{\sigma }(\omega )$
describes the tunneling of spin-$\sigma $ electrons across the junction and
is given by\ $\mathcal{T}_{\sigma }(\omega )=\frac{4\Gamma _{L}\Gamma _{R}}{%
\Gamma _{L}+\Gamma _{R}}A _{\sigma }(\omega )$ where $A _{\sigma
}(\omega )$ is the spin dependent spectral density of the central QD \cite{cornaglia2005strongly,cornaglia2008comment}.
In the above expression $\Gamma _{\alpha }=\pi\sum_k |V_{k\alpha }|^{2}\delta(\epsilon_{k\alpha})
$ is the contribution to the width of the molecular energy levels introduced
by the coupling with lead $\alpha$, and $f(\omega )$ is the Fermi function. In what follows we assume that $\Gamma_L$ and $\Gamma_R$ are equal and energy independent~\footnote{For a discussion about this assumptions see e.g. Ref. \cite{Costi2010}.}, define the effective tunnel coupling $V=\sqrt{(\Gamma_L+\Gamma_R)/\pi \rho}$, where $\rho=1/(2D)$ is the local density of states of the electrodes at the Fermi energy ($\varepsilon_F=0$), and choose half the bandwidth of the leads $D$ as the unit of energy. 
The total charge current is $I_c=\sum_\sigma I_{\sigma}$ and we define the pure spin current for $I_c=0$ as
\begin{equation}
	I_s = S_s\frac{\Delta T}{|e|\mathcal{R}},
	\label{ieq:spincurr}
\end{equation}
with 
\begin{equation}
	S_s=S_\uparrow-S_\downarrow,
	\label{Ss}
\end{equation}
the spin Seebeck coefficient and $\mathcal{R}=\sum_\sigma G_\sigma^{-1}$.

In what follows we present results for the transport properties calculating $A_\sigma(\omega)$ using Wilson's Numerical Renormalization Group \cite{bulla2008numerical}.

\subsection{Conductance}
We first analyze the behavior of the conductance as a function of the temperature for different values of the interdot hopping amplitude $t$. 
The system presents, as a function of $t$, a crossover from a ferromagnetic Kondo regime, for the larger $t$ values, to a two stage Kondo 
regime in the low $t$ range of values.  

\begin{figure}[tbp]
	\includegraphics[width=\columnwidth]{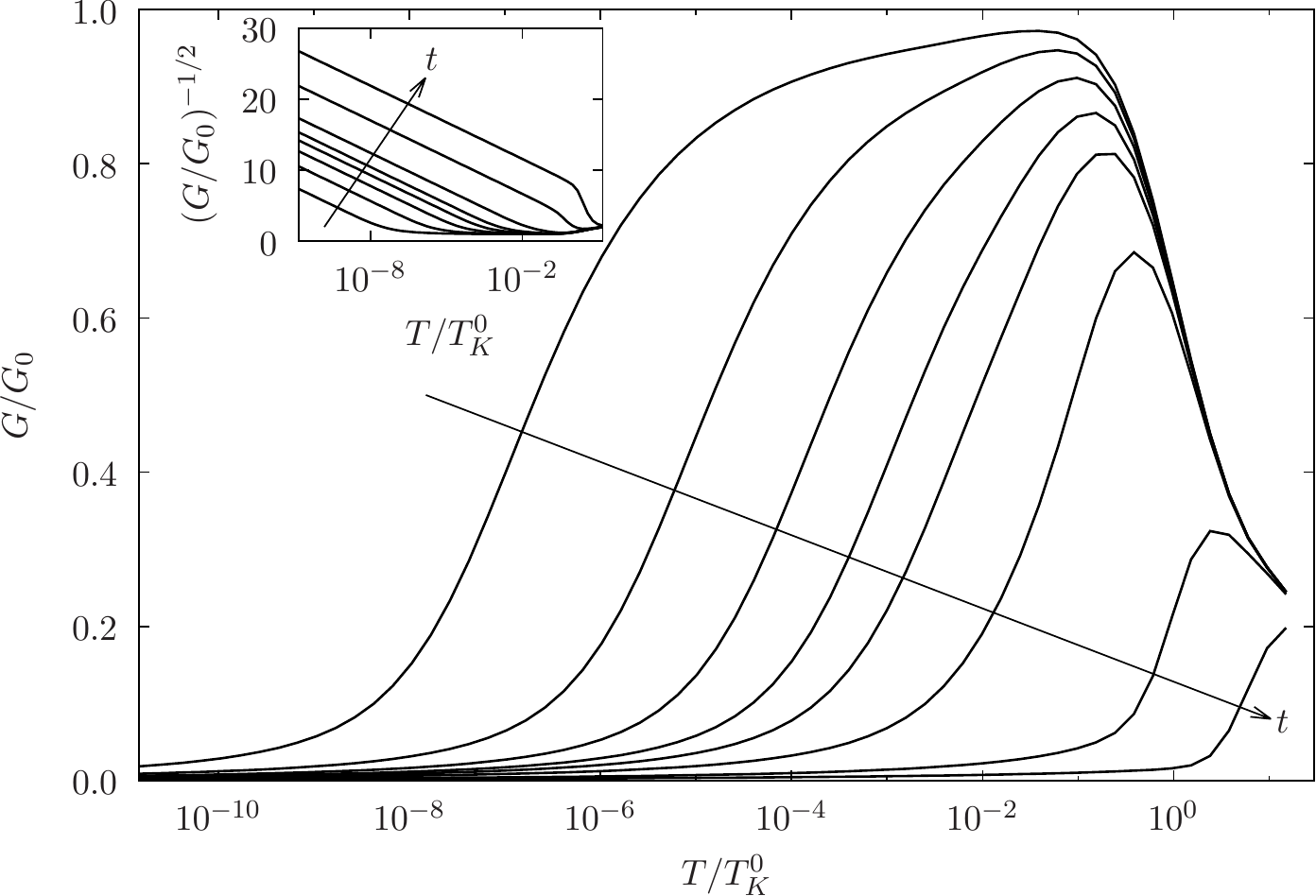}
	\caption{(Color online) Conductance as function of the temperature for different values of the tunnel coupling $t$. $t/D$ takes the values $0.012$, $0.014$, $0.016$, $0.018$, $0.02$, $0.025$, $0.05$ and $0.1$. Other parameters are $U=0.4D$, $V=0.2D$ and 
$\epsilon_\ell=-U/2$. Inset: plot of $G^{-1/2}$ as a function of $\ln(k_BT/T_K^0)$ to make clear the singular behavior of the conductance 
at low energies for $t\neq0$.} 
\label{fig:GvsTt}
\end{figure}
Figure \ref{fig:GvsTt} presents the conductance through a three QD device in the electron-hole symmetric situation ($\varepsilon_\ell=-U/2$) as a function of the temperature for different values of the interdot hopping amplitude $t$. 
In the lower $t$ regime, there is a increase of the conductance as the temperature is lowered and it reaches a value $G\sim G_0=2e^2/h$ for $T\lesssim T_K^0$, where $T_K^0$ is the Kondo temperature for the central QD. 
At temperatures below a characteristic temperature $T_0$, the underscreened Kondo effect sets in and the conductance decreases again to $G\sim 0$ for $T\ll T_0$. 
As $t$ is increased, $T_0$ increases, and the high conductance plateau is suppressed. 
In the larger $t$ regime, no high conductance plateau is obtained and the conductance decreases monotonously. 
In this case the system can be described by the ferromagnetic Kondo model. In the whole range of values of $t$ studied, the NRG results 
for the low temperature conductance present a singular behavior of the form
\begin{equation}
	G(T\to 0)\simeq G(T=0)+\frac{b_G}{\ln^2(T/T_0)}
	\label{eq:lowTG}
\end{equation}
where $k_BT_0$ is much larger than $D$ in the ferromagnetic Kondo regime and much smaller than $D$ in the underscreened Kondo regime \cite{andrade2015ferro}.
For the parameters of Fig. \ref{fig:GvsTt}, $G(T=0)=0$, while assuming a regular Fermi liquid ground state $G(T=0)=G_0$ would be expected \cite{andrade2014transport}:
\begin{equation}
	G^{FL}(T=0) = \frac{G_0}{2}\sum_\sigma \sin^2(\mathcal{N}_{\sigma}\pi),
	\label{eq:GFL}
\end{equation}
where $\mathcal{N}_\sigma=3/2$ is the average occupancy per spin. A qualitatively identical behavior is obtained for systems having a 
larger number of side coupled QDs, the main difference being the value of $T_0$~\cite{andrade2015ferro}. 
\begin{figure}[tbp]
\includegraphics[width=\columnwidth]{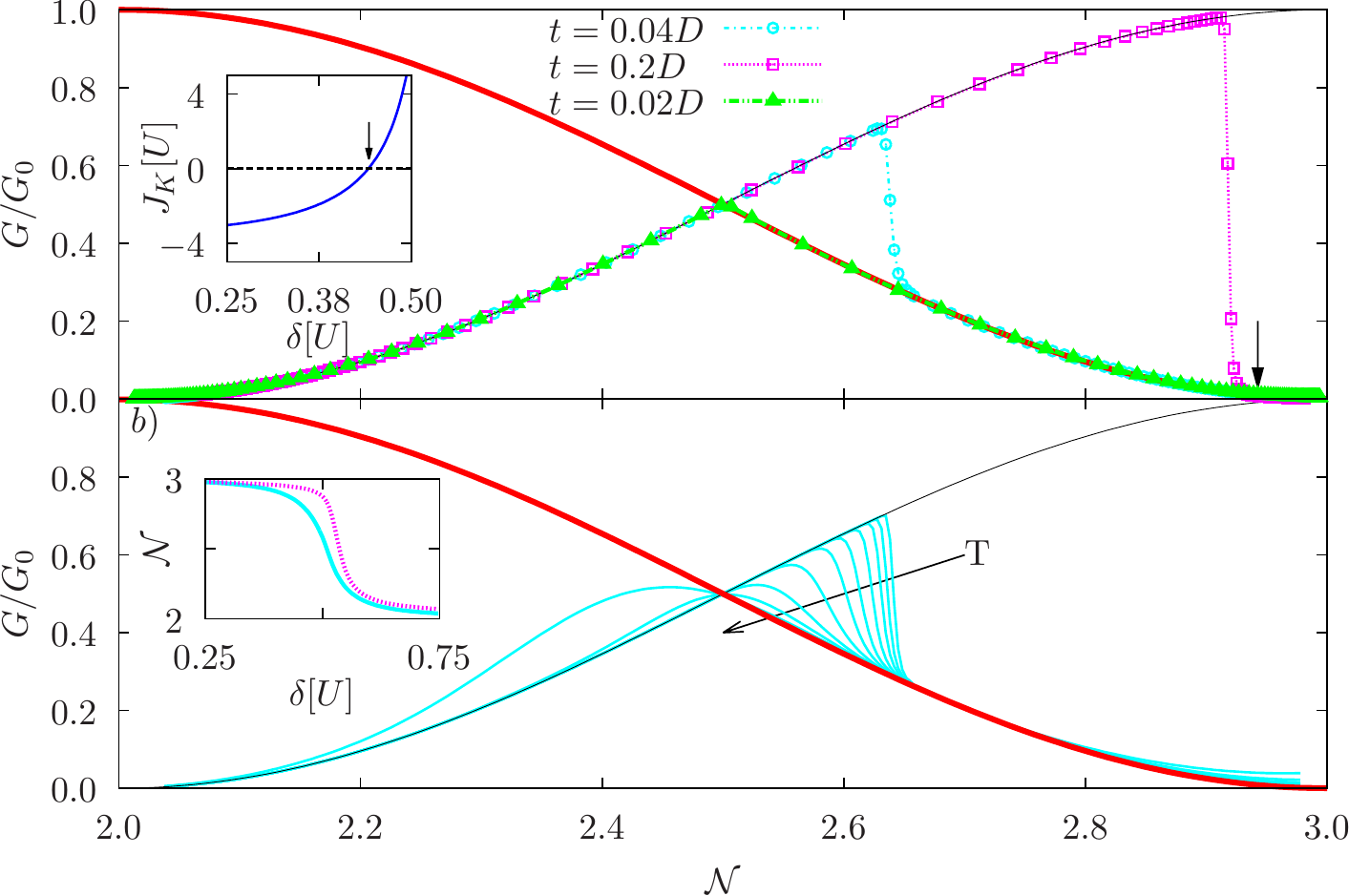}
\caption{(Color online) Conductance $G$ vs total occupation $\mathcal{N}$ of the QDs. a) Conductance for different couplings $t$. Inset: Kondo coupling as a function of $\delta$ for $t=0.2D$. 
The arrow indicates the value of $\delta=\delta^\star\simeq 0.44U$ where $J_K$ vanishes. b) Conductance for $t=0.04D$ at different temperatures. $k_BT$ changes from $10^{-13}D$ to $10^{-4}D$ 
in steps of $1$ in the exponent. Inset: $\mathcal{N}$ as a function of $\delta$ for $t=0.04D$ (solid line) and $t=0.2D$ (dashed line). In a) and b) the expected zero-temperature conductance for a singular Fermi liquid $G_0 \sin^2(\pi \mathcal{N}/2-\pi/2)$ (red thick line) and a regular Fermi liquid  $G_0\sin^2(\pi \mathcal{N}/2)$ (black thin line), are presented.}
\label{fig:QCPdelta}
\end{figure}

Figure \ref{fig:QCPdelta} presents the low temperature conductance as a function of total occupation $\mathcal{N}$, in the QD array, which is changed by shifting the energy of one of the side coupled QDs by $\delta$ ($\varepsilon_{0}=\varepsilon_{1}=-U/2$ and $\varepsilon_2=-U/2+\delta$). For $\mathcal{N}=3$ the system is in a singular Fermi liquid regime and the conductance vanishes. As $\mathcal{N}$ is reduced, the conductance follows the behavior that would be expected for a Fermi liquid with an effective charge per spin projection sector in the QD array reduced by $\pi/2$: 
\begin{equation}
	G(T\to 0)=	G_0 \sum_\sigma \sin^2[(\mathcal{N}_\sigma-1/2)\pi].
	\label{eq:GNFL}
\end{equation}
This can be interpreted as due to the asymptotic decoupling at low energies of a spin $1/2$ from the reservoir, reducing the effective charge in the QD array. For values of $t$ such that at $\delta=0$ ($\mathcal{N}=3$) the system is in the ferromagnetic Kondo regime, the zero temperature conductance presents a discontinuity as a function of $\mathcal{N}$. 
A calculation of the Kondo coupling between the QD array spin and the reservoirs shows that there is a change in the sign of $J_K$ at the value of $\delta$ where the jump in the conductance is observed. The inset of Fig. \ref{fig:QCPdelta}a) shows that the change in the sign of $J_K$ for $t=0.2D$ takes place at $\delta=\delta^\star\simeq0.44U$ which corresponds to a total occupation of $\mathcal{N}=2.942$ [see inset of the Fig. \ref{fig:QCPdelta}b)] marked by the arrow in the Fig. \ref{fig:QCPdelta}a).
As $J_K$ decreases in absolute value, the Kondo temperature vanishes with an essential singularity behavior and a Kosterlitz-Thouless transition for $J_K=0$ is obtained where the system changes from being a singular Fermi liquid to a regular Fermi liquid. This changes the zero temperature behavior of the conductance which is described by Eq. (\ref{eq:GFL}) for $J_K>0$ and by Eq. (\ref{eq:GNFL}) for $J_K<0$.
Close to the transition, the temperature needs to be very low (lower than $T_K$) for the conductance to attain its zero temperature value. At finite temperatures the discontinuity in $G(\mathcal{N})$ is rounded as can be seen in Fig. \ref{fig:QCPdelta} b), although a strong gate voltage dependence of the conductance remains close to the transition.
For the lower values of $t$ analyzed, the system is in the underscreened Kondo regime, the transition occurs close to $\mathcal{N} = 2.5$ and the conductance displays a cusp at low temperatures.

\subsection{Thermopower}
\begin{figure}[tbp]
\includegraphics[width=\columnwidth]{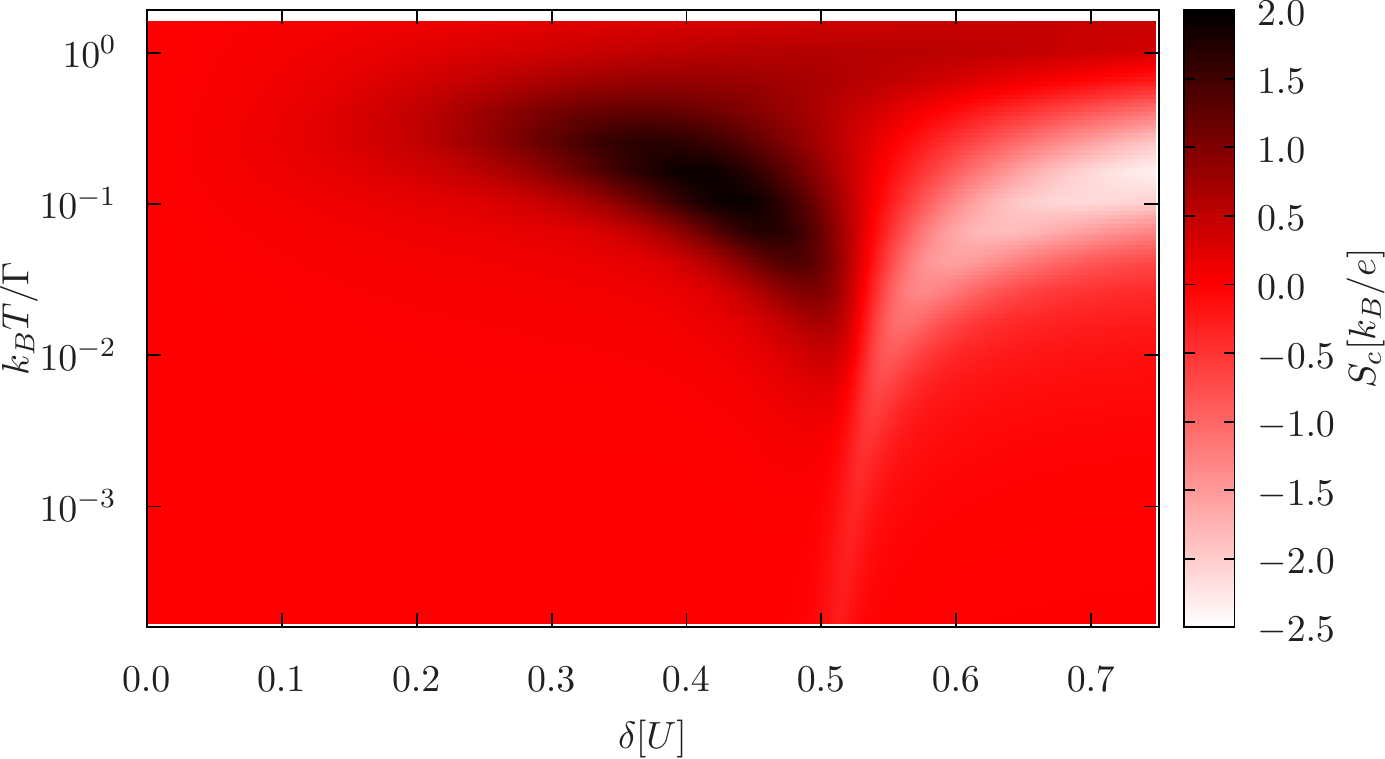}
\caption{(Color online) Charge Seebeck coefficient $S_c$ in the ferromagnetic Kondo regime ($t=0.2D$) for $B=0$. There is a change in the sign of $S$ near $\delta\sim  U/2$ where the total occupation of the QDs changes between $\sim 3$ and $\sim 2$. Other parameters as in Fig. \ref{fig:GvsTt}.}
\label{fig:chargeS}
\end{figure}
Figure \ref{fig:chargeS} presents an intensity map of the charge Seebeck coefficient as a function of the temperature and the energy shift $\delta$ 
for a system which, for $\delta=0$, is in the ferromagnetic Kondo regime ($t=0.2D$). In the electron-hole symmetric situation 
($\delta=0$), electrons and holes contribute the same to the Seebeck effect but with opposing signs leading to a zero Seebeck 
coefficient in the full range of temperatures. A finite $\delta$ breaks the electron-hole symmetry and leads to a finite charge 
Seebeck coefficient which has a peak as a function of the temperature at $T\sim 0.1\Gamma$. 
For $\delta\sim U/2$ where the occupation on the QD array changes between $\sim3$ and $\sim2$, there is a change in the sign 
of $S_c$ and a sharp feature in $S_c$ remains even at the lowest temperatures studied.
For negative values of $\delta$, the sign of $S_c$ is inverted $S_c(-\delta)=-S_c(\delta)$ due to the electron-hole replacement symmetry.
\begin{figure}[tbp]
\includegraphics[width=\columnwidth]{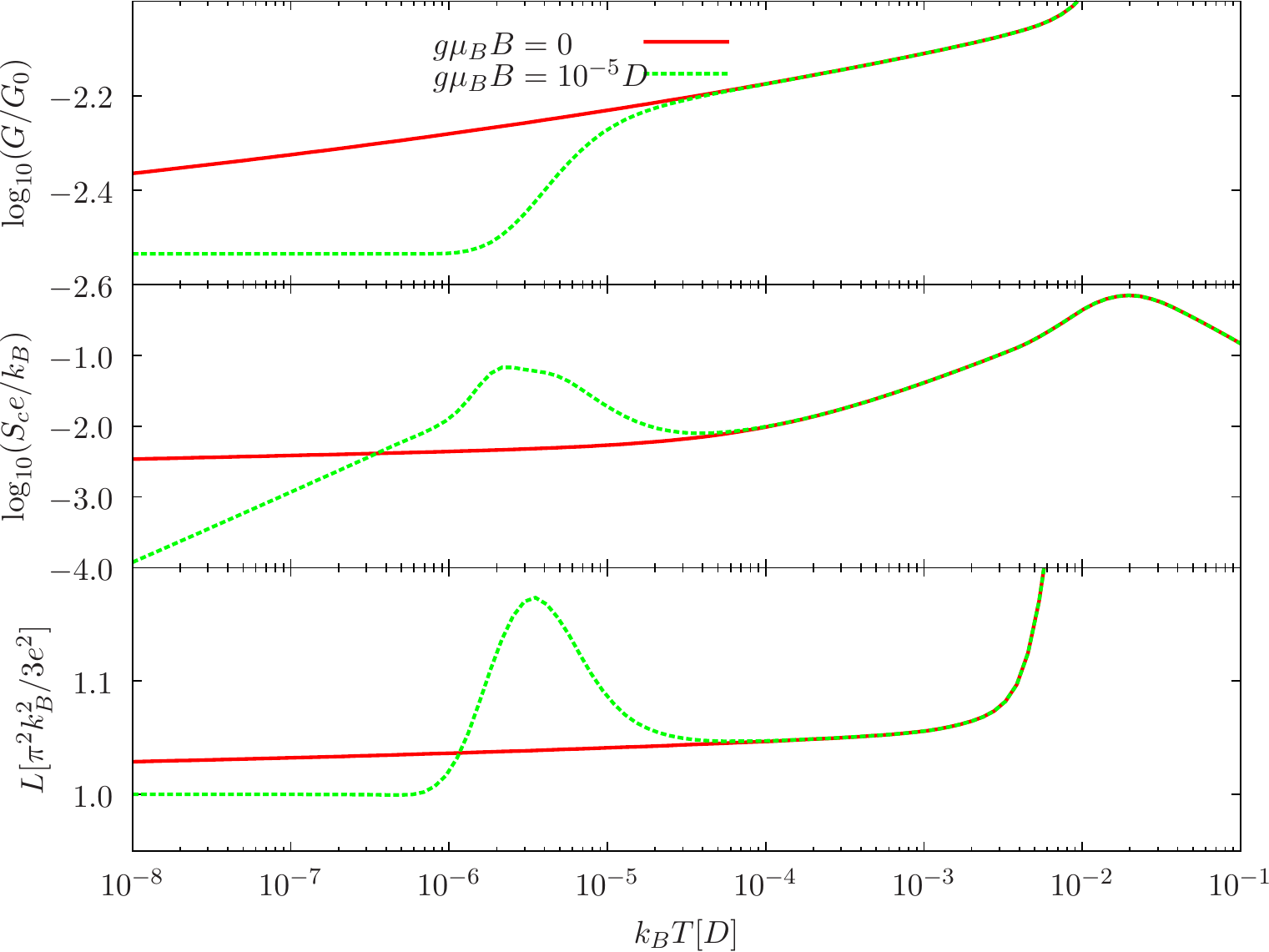}
\caption{(Color online) From top to bottom panel: Conductance, Seebeck coefficient $S_c$, and Lorentz number, as a function of the 
temperature for a system in the ferromagnetic Kondo regime $t=0.2$ and $\delta=0.23U$. Other parameters as in Fig. \ref{fig:GvsTt}.}
\label{fig:GSL}
\end{figure}

A Sommerfeld expansion at low temperatures for the integrals of Eq. (\ref{GandS}) leads to
\begin{equation}
	S_\sigma(T)=-\frac{k_B}{|e|}\frac{\pi^2}{3}k_B T\frac{1}{A_\sigma(\omega=0,T)}\left.\frac{\partial A_\sigma(\omega,T)}{\partial \omega}\right|_{\omega=0}.
	\label{eq:lowTS}
\end{equation}
and a linear in $T$ vanishing of the Seebeck coefficient is expected for a Fermi liquid as $T\to 0$.
The spectral density in this system in the singular Fermi liquid regime has, at low temperatures ($T \to 0$), a behavior of the form \cite{andrade2015ferro}
\begin{equation}
	A_\sigma(\omega\to 0)\simeq  \frac{b}{\ln^2(|\omega|/k_B T_0)},
	\label{singAlow}
\end{equation}
and Eq. (\ref{eq:lowTS}) is not applicable. The charge Seebeck coefficient does vanish at low temperatures but slower than linear 
in $T$ due to logarithmic corrections. This can be observed in Fig. \ref{fig:GSL}, where the conductance and the Lorenz number also 
show a logarithmic temperature dependence at low temperatures. This singular behavior is cutoff by an external magnetic field that 
drives the system into a Fermi liquid regime making Eq. (\ref{eq:lowTS}) with a linear behavior in $T$ of $S_c$ applicable. 

\subsection{Spin Seebeck effect and spin polarization}
\begin{figure}[tbp]
\includegraphics[width=\columnwidth]{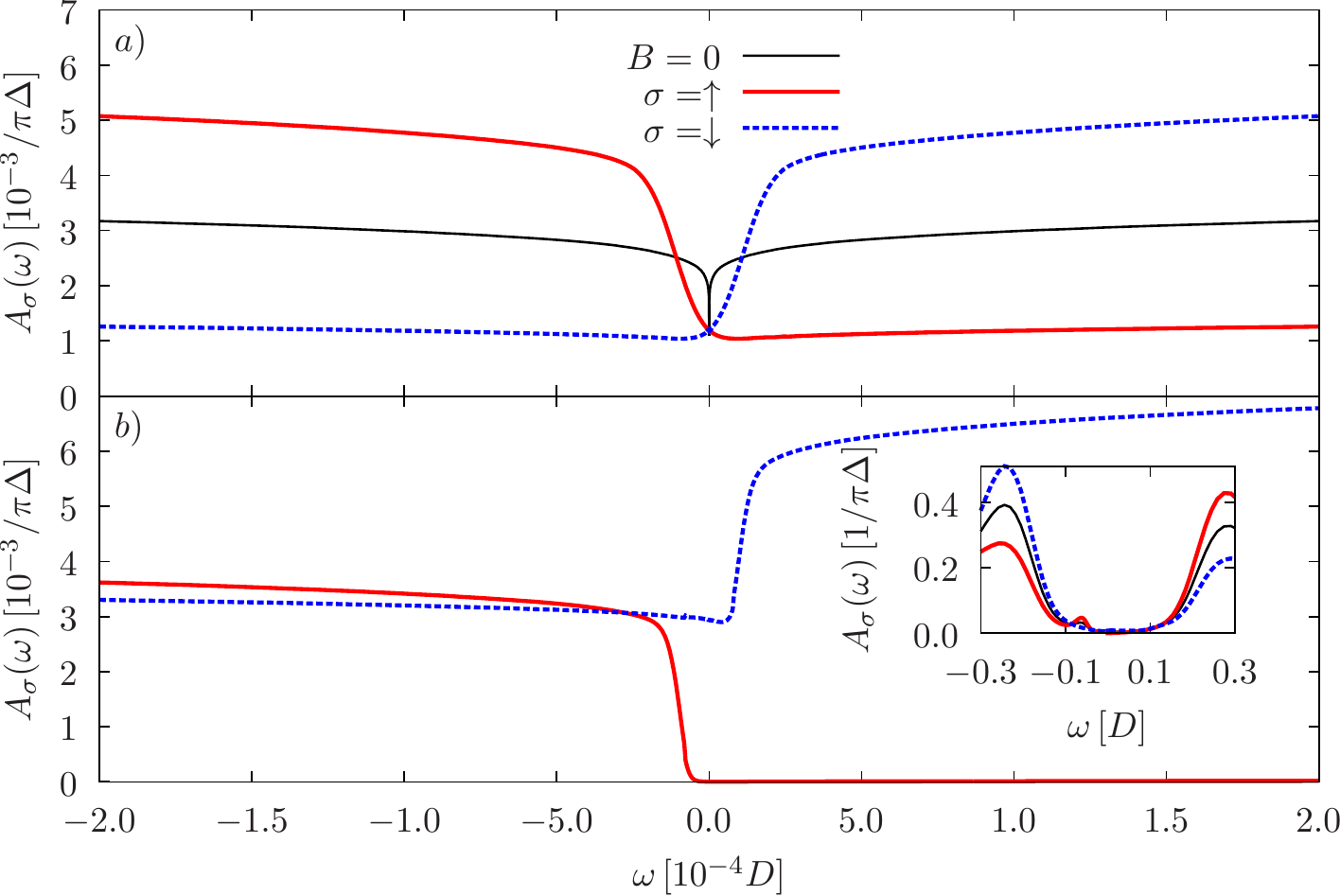}
\caption{(Color online) Spectral density of the central QD at low energies 
for $k_BT = 0.3\mu_B B=3\times 10^{-6}D$ in the ferromagnetic 
regime ($t=0.2D$) and $\delta=0$. For $B=0$ and the spectral density is electron-hole symmetric and presents a singular behavior at low energies. 
$b)$ Same as a) for $\delta=0.23U$ and $k_BT=10^{-6}D$. Inset: Spectral density in a wide range of energies that includes the Hubbard peaks at $\omega \sim \pm U/2$ and shows the spin polarization under an applied magnetic field.}
\label{AwVp02Txkdx}
\end{figure}
\begin{figure}[tbp]
\includegraphics[width=\columnwidth]{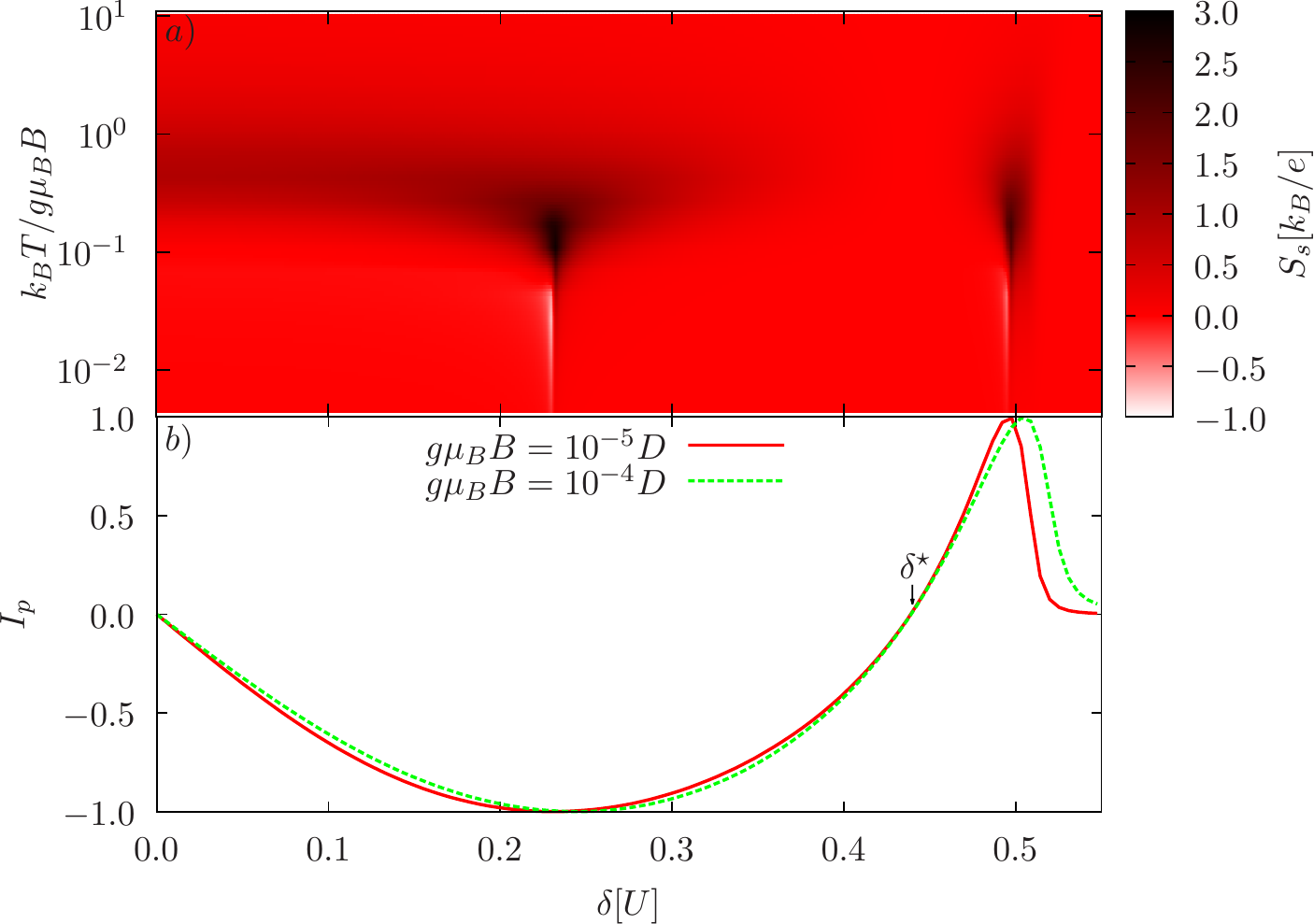}
\caption{(Color online) a) Spin Seebeck coefficient in the ferromagnetic Kondo regime ($t=0.2D$) and $g\mu_BB=10^{-5}D$. There are two peaks associated with 
with a maximum of $S_\uparrow$ at $\delta=0.23U$ and a minimum of $S_\downarrow$ at $\delta\sim U/2$.
b) Zero temperature current polarization $I_p$ as a function of $\delta$. There is a broad range of values of $\delta$ around $\delta\sim 0.23U$ and  where the current is $\sim 100\%$ polarized.
}
\label{fig:mapSs}
\end{figure}

In the absence of an external magnetic field the system preserves the spin symmetry, $S_\uparrow\equiv S_\downarrow$, and the spin Seebeck coefficient is zero. 
In the electron-hole symmetric situation ($\delta=0$), $A_\uparrow(\omega)=A_\downarrow(-\omega)$ [see Fig. \ref{AwVp02Txkdx}a)] and $S_\uparrow=-S_\downarrow$ which leads to a finite $S_s$ under a finite magnetic field $B$.  Figure \ref{fig:mapSs}a) presents the spin Seebeck coefficient as a function of the temperature and the 
energy shift $\delta$. 
For $\delta<\delta^\star$ the system is in a singular Fermi liquid regime and the spin is easily polarized at low temperatures by an external 
magnetic field. This leads to a strong electron-hole asymmetry for $A_\sigma(\omega)$ and to a large $S_s \sim k_B/|e|$ for $k_B T\sim g\mu_B B$.
For $k_B T \ll g \mu_B B$, Fermi liquid theory is applicable and $S_s$ is proportional to the temperature.

For both $\delta\sim 0.23U$ and $\delta \sim U/2$, $S_s$ presents a large peak and a sharp feature with a change of sign at low temperatures. These features are associated with the vanishing of $G_\uparrow$ and $G_\downarrow$, respectively, which leads to a spin polarization of the current ($\sim 100\%$). This can be observed in Fig. \ref{fig:mapSs}b) which presents the current polarization factor $I_P = \frac{G_\uparrow-G_\downarrow}{G_\uparrow+G_\downarrow}$ as a function of $\delta$. 
$|I_P|$ attains its maximum values where the spin Seebeck coefficient is maximum. For $k_B T\ll g\mu_B B$ the system is in a Fermi liquid regime and Eq. (\ref{eq:GFL}) for the conductance is valid. 
The total occupation and the magnetization are changed by the energy shift $\delta$ and the magnetic field, and we have $\mathcal{N}=\mathcal{N}_\uparrow+\mathcal{N}_\downarrow=3-\Delta \mathcal{N}$ and 
$M=\mathcal{N}_\uparrow-\mathcal{N}_\downarrow=1-\Delta M$, where $0\leq \Delta \mathcal{N}\leq 1$ for all $\delta\geq0$, and 
$\Delta M\geq0$ for $0\leq\delta\leq\delta^{\star}$ and $\Delta M<0$ for $\delta>\delta^{\star}$. From the total occupation and magnetization
we get the spin occupation as
\begin{equation}
	\mathcal{N}_\uparrow=2+(\Delta M-\Delta \mathcal{N})/2,
	\label{N_up}
\end{equation}
\begin{equation}
	\mathcal{N}_\downarrow=1-(\Delta M+\Delta \mathcal{N})/2.
	\label{N_dw}
\end{equation}
For $\delta=0.23U$, $\Delta M=\Delta \mathcal{N}$ and $\mathcal{N}_\uparrow$ is an integer number that result in $G_\uparrow=0$ 
[see Eq. (\ref{eq:GFL})], while for $\delta\sim U/2$, $\Delta M=-\Delta \mathcal{N}$, $\mathcal{N}_\downarrow$ is an integer 
number and $G_\downarrow=0$. The vanishing of the conductance for one of the spin projections at low temperatures, is associated 
with the vanishing of the spectral density (see Fig. \ref{AwVp02Txkdx}).

\begin{figure}[tbp]
\includegraphics[width=\columnwidth]{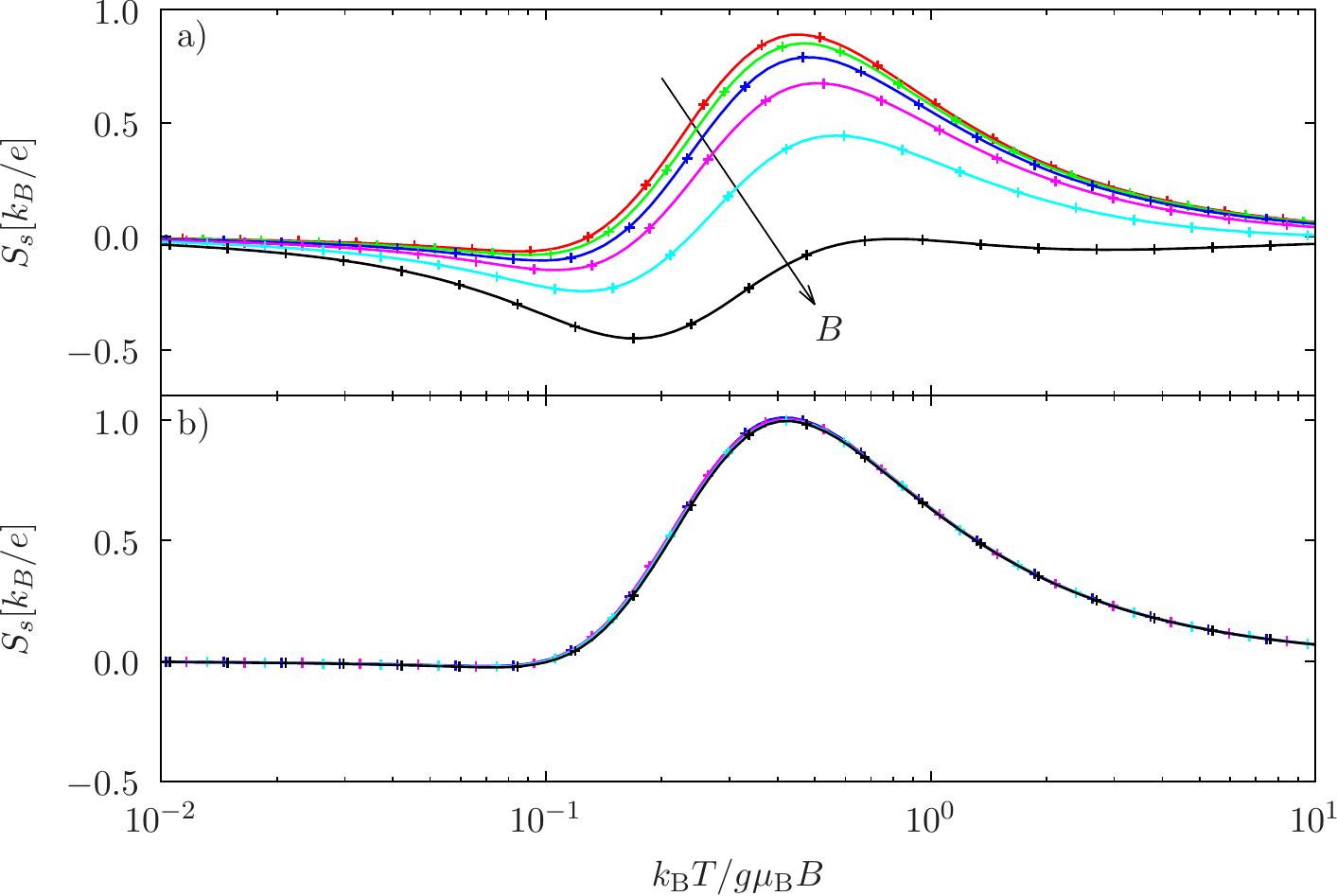}
\caption{(Color online) Spin Seebeck as function of the temperature and the magnetic field at $\delta=0$ for a) $t=0.02D$ and b) $t=0.2D$. 
The magnetic field changes from $10^{-11}D$ to $10^{-6}D$ in steps of $1$ in the exponent.}
\label{fig:SSeb}
\end{figure}

Figure \ref{fig:SSeb} presents the spin Seebeck coefficient as a function of the temperature scaled by the applied magnetic field 
for an electron-hole symmetric situation ($\delta=0$). 
We find a large peak in $S_s$ for $k_BT\sim g\mu_BB$  that loses strength when the magnetic field increases in 
the underscreened Kondo regime [see Fig. \ref{fig:SSeb}a)], and holds with essentially the same strength for the 
ferromagnetic Kondo regime [see Fig. \ref{fig:SSeb}b)]. This is due to the fact that in the underscreened Kondo regime the energy 
scale $k_BT_0$, that determines the onset for the singular behavior, is much lower than the bandwidth \cite{andrade2015ferro} and 
an external magnetic field easily can drive the system out the underscreened Kondo regime. In the ferromagnetic Kondo regime, however, 
the energy scale $k_BT_0$ is much larger than the bandwidth \cite{andrade2015ferro} and the onset of the singular behavior occurs at 
a higher energy scale ($\sim U$).

\section{Summary and Conclusions}\label{sec:sumconcl}

We have analyzed the thermoelectric and magnetoelectric properties of a three QDs device
in a star configuration as a function of temperature, magnetic field and gate voltage. 
The system presents a high sensitivity to external fields, associated with quantum phase transitions, that manifests on the transport properties.
The zero-temperature conductance as a function of gate voltage presents a discontinuity that signals a transition between singular and regular Fermi liquid regimes. This quantum phase transition is produced by a change in the sign of the Kondo coupling $J_K$ between a magnetic moment in the QD array and the leads. 
For an antiferromagnetic coupling $J_K>0$ the low temperature properties of the system can be described with Fermi liquid theory, and the zero-temperature conductance is given by the QD array occupation through Friedel's sum rule. For a ferromagnetic coupling $J_K<0$ the system in a singular Fermi liquid regime at low temperatures, and satisfies a modified sum rule where the total charge in the QD array is replaced by a reduced effective charge. The effective charge can be obtained by subtracting the charge associated with a magnetic moment on the QD array that decouples asymptotically from the leads. 
The conductance, thermopower, and Lorentz number present a logarithmic behavior at low temperatures in the ferromagnetic Kondo regime. An external magnetic field strongly polarizes the asymptotically free magnetic moment and drives the system into a Fermi liquid regime, producing a low energy cutoff for the logarithmic behavior of the transport properties. In this regime, the magnetic field produces a large slope at low energies in the spectral density of the central QD, with a sign that depends on the spin projection along the external magnetic field. Close to the electron-hole symmetric condition this leads to a large spin Seebeck coefficient ($\sim k_B/|e|$) and to pure spin currents for $k_B T\sim g\mu_B B$ in a wide range of magnetic fields. A gate voltage can tune the system to produce spin polarized currents due to the suppression of the current of one of the spin projections caused by interference effects. 

For a weak tunnel coupling between QDs, the system presents a two-stage Kondo regime with an antiferromagnetic Kondo effect for a spin 1/2 followed by a spin-1 underscreened Kondo effect. In this case, the electronic properties of the system also present a singular behavior but it is attained at much lower temperatures. 

We analyzed devices with more than three QDs ($N>2$), where $N$ QDs are tunnel coupled to the central QD, and found a qualitatively identical behavior to the $N=2$ case for the thermoelectric and magnetoelectric properties. The main quantitative differences are a reduced maximal spin Seebeck coefficient and an enhanced energy scale $k_BT_0$ as the number of side-coupled QDs $N$ increases.

\acknowledgments
This work was partially supported by CONICET PIP0832, SeCTyP-UNCuyo 06C347, and PICT 2012-1069.
\bibliographystyle{apsrev4-1}
\bibliography{references}

\end{document}